# Real-space Observation of Unidirectional Charge Density Wave and Complex Structural Modulation in the Pnictide Superconductor $Ba_{1-x}Sr_xNi_2As_2$

Tian Qin[1,*], Ruixia Zhong[1,*], Weizheng Cao[1,*], Shiwei Shen[1], Chenhaoping Wen[1,†],

Yanpeng Qi[1,2,3,†], and Shichao Yan[1,2,†]

[1] *School of Physical Science and Technology, ShanghaiTech University, Shanghai 201210, China*

[2] *ShanghaiTech Laboratory for Topological Physics, ShanghaiTech University, Shanghai 201210, China*

[3] *Shanghai Key Laboratory of High-resolution Electron Microscopy, ShanghaiTech University, Shanghai 201210, China*

[*] *These authors contributed equally to this work*

[†] *Email: wenchhp@shanghaitech.edu.cn; qiyp@shanghaitech.edu.cn yanshch@shanghaitech.edu.cn*

## ABSTRACT

Here we use low-temperature and variable-temperature scanning tunneling microscopy to study the pnictide superconductor, $Ba_{1-x}Sr_xNi_2As_2$. In the low-temperature phase (triclinic phase) of $BaNi_2As_2$, we observe the unidirectional charge density wave (CDW) with $Q = 1/3$ on both the Ba and NiAs surfaces. On the NiAs surface of the triclinic $BaNi_2As_2$, there are structural-modulation-induced chain-like superstructures with distinct periodicities. In the high-temperature phase (tetragonal phase) of $BaNi_2As_2$, the NiAs surface appears as the periodic $1\times2$ superstructure. Interestingly, in the triclinic phase of $Ba_{0.5}Sr_{0.5}Ni_2As_2$, the unidirectional CDW is suppressed on both the Ba/Sr and NiAs surfaces, and the Sr substitution stabilizes the periodic $1\times2$ superstructure on the NiAs surface, which enhance the superconductivity in $Ba_{0.5}Sr_{0.5}Ni_2As_2$. Our results provide important microscopic insights for the interplay among the unidirectional CDW, structural modulation, and superconductivity in this class of pnictide superconductors.

Unconventional superconductivity is often accompanied by other novel electronic orders, such as charge density wave (CDW), spin density wave (SDW), and nematic orders. Understanding the interplay between superconductivity and the intertwined orders remains one of the major issues in the research of quantum materials.[1-3] CDW in cuprates and nematic fluctuations in iron-based superconductors have been intensively studied,[4-8] and CDW and nematicity have also been observed in the superconducting magic-angle graphene.[3] Although it has been widely believed that these electronic orders have strong interrelation with superconductivity, their precise role in superconductivity has not been fully understood. Recently, several CDW states and the nematic order have been reported in the $BaNi_2As_2$-related superconducting materials,[9-17] which makes them a new platform for exploring the interplay between superconductivity and the intertwined orders.

$BaNi_2As_2$ is a nickel-based and nonmagnetic pnictide superconductor ($T_c \sim 0.7$ K).[18-21] At room temperature, it shares the same tetragonal phase as the 122-family Fe-based superconductor, $BaFe_2As_2$.[18, 22, 23] Upon cooling, $BaNi_2As_2$ first enters the incommensurate CDW state with the wavevector $Q = 0.28$ below ~150 K, then undergoes the first-order structural transition from the tetragonal to the triclinic crystal structure at ~136 K. In the triclinic phase, the system exhibits a slightly incommensurate CDW with $Q = 0.31$, then locks into a unidirectional commensurate CDW with $Q = 1/3$ by further cooling.[9, 10] The very recent angle-resolved photoemission spectroscopy (ARPES) results also show the spectral evidence for the unidirectional-CDW-induced band folding in $BaNi_2As_2$.[13] Near the triclinic phase transition, the elastoresistance measurements in $BaNi_2As_2$ indicate the diverging of the nematic susceptibility in the $B_{1g}$ channel, and the Sr substitution can gradually suppress the nematic order, which may be the reason for the six-fold increasement of the superconducting $T_c$ in the $Ba_{1-x}Sr_xNi_2As_2$ system.[11] As increasing the Sr substitution, the triclinic phase transition temperature decreases, and a new CDW with $Q = 1/2$ emerges.[10] Until now, for the $Ba_{1-x}Sr_xNi_2As_2$ superconductor, the microscopic information about the structure, the unidirectional CDW, and the nematicity remains unknown.

In this work, we report a low-temperature and variable-temperature scanning tunneling microscopy/spectroscopy (STM/STS) study on $Ba_{1-x}Sr_xNi_2As_2$. In the triclinic phase of $BaNi_2As_2$, we find the unidirectional CDW on both the Ba- and the NiAs-terminated surfaces. On the NiAs surface of the triclinic $BaNi_2As_2$, we observe the chain-like complex structures with several distinct periodicities. In the tetragonal $BaNi_2As_2$, there is a periodic $1 \times 2$ superstructure on the NiAs surface. We also study the influence of Sr substitution on the unidirectional CDW, structure and superconductivity in $Ba_{1-x}Sr_xNi_2As_2$. Our study provides detailed microscopic information about unidirectional CDW and structural modulation in $Ba_{1-x}Sr_xNi_2As_2$, which is essential for understanding the nematicity and superconductivity in this class of superconductors.

The tetragonal to triclinic phase transition in $BaNi_2As_2$ happens at ~136 K, and we first investigate the triclinic $BaNi_2As_2$ at 4.2 K. Cleaving the $BaNi_2As_2$ single crystals results in both the Ba- and NiAs-terminated surfaces. Figure 1c shows the constant-current STM topography taken on the Ba surface of $BaNi_2As_2$, where we can clearly see the stripe-like unidirectional pattern. The inset in Figure 1c shows its Fourier transform (FT) image, which indicates that the periodicity of the stripe-like pattern is three times the length of the lattice constant. The typical differential conductance ($dI/dV$) spectra taken on and off stripe marked in Figure 1c have clear differences at the energies above ~150 mV (Figure 1d). The appearance of the stripe modulation has a phase shift in the STM topographies taken with the bias voltages below and above ~150 mV (Figure 1e), which is a key feature for a CDW (see Figure S2). Our STM data indicates that there is indeed a unidirectional commensurate CDW with the wavevector $Q = 1/3$ on the Ba surface. In addition to the unidirectional CDW, we can also see the ring-like standing wave patterns around the atomic defects (Figure 1c), which is due to the quasiparticle interference (QPI) of the electronic states.

In order to extract the electronic band dispersion from the QPI, we perform the energy-resolved differential conductance ($dI/dV$) maps. Figure 1f shows the typical $dI/dV$ map taken with −30 mV bias voltage (see Figure S3). The QPI pattern can be obtained by performing FT to the $dI/dV$ maps. Figure 1g is the FT image of the $dI/dV$ map shown in Figure 1f, where there are three main sets of inequivalent scattering wavevectors: $q_1$, $q_2$ and $q_3$. As the first approximation for understanding these scattering wavevectors, we perform an autocorrelation of the constant energy contours, which neglects the spin or orbital matrix elements during the scattering.[24] Considering the size of the scattering wavevectors in Figure 1g, we only plot the autocorrelation of the four small Fermi pockets around $\Gamma$ point in the constant energy contours (the red circles shown in Figure 1b).[25] The autocorrelation pattern can nicely reproduce the main scattering modes shown in Figure 1g (see Figure S4). The $q_1$ and $q_3$ wavevectors indicate the scattering along the $\Gamma$-X and $\Gamma$-Y directions, respectively. The $q_2$ wavevector corresponds to the scattering along the $\Gamma$-M direction.

By plotting the energy dispersion of the $q_1$ and $q_2$ wavevectors, we observe an electron-like band crossing the Fermi level with the band bottom at $\sim-150$ mV (Figure 1h), which is consistent with the $\alpha$ band in the previous ARPES measurements.[25] In the d$I$/d$V$ spectrum (Figure 1d), the bottom of the $\alpha$ band appears as a peak located at $\sim-150$ mV. As marked by the green arrow, we can also see the nondispersive unidirectional CDW wavevector in Figure 1h.

We next turn to the NiAs surface of the triclinic BaNi$_2$As$_2$. We observe two different NiAs surfaces, and Figure 2a shows the STM topography taken near the boundary of these two NiAs surfaces (Domain-I and Domain-II). Figures 2b and 2c are the zoom-in STM topographies taken on these two surfaces, respectively. We can see the chain-like modulation patterns with distinct periodicities on both Domain-I and Domain-II NiAs surfaces, and the chain directions on these two domains are perpendicular with each other (Figures 2b and 2c). According to the very recent first-principles calculations, the rebonding of the As anions in the NiAs layer of BaNi$_2$As$_2$ can induce complex structural instabilities, such as As dimers, trimers or even more complex bonding arrangements.[26] These structural instabilities are close in energy and can coexist in BaNi$_2$As$_2$, which makes the NiAs surface appear as the complex superstructures with several different periodicities. Similar chain-like complex superstructures have also been reported in the triclinic IrTe$_2$ which is due to the rebonding of Te.[27]

As shown in Figure 2c, in addition to the chain-like superstructures, the unidirectional CDW pattern can be clearly seen on the Domain-II NiAs surface, and it is perpendicular with the chain-like modulation pattern. In order to clearly reveal the periodicities of the chain-like modulations and the unidirectional CDW on these two NiAs surfaces, we perform FT to the STM topographies (Figures 2d and 2e). Figures 2f and 2g are the linecut profiles taken along the colored arrows in Figures 2d and 2e, respectively. From the red curve in Figure 2g, we can see the strong wavevector corresponding to the $Q = 1/3$ unidirectional CDW on the Domain-II surface. As shown by the red curve in Figure 2f, the chain-like superstructures indeed consist of several periodicities, and the $1\times3$ wavevector can also be seen. This indicates that the unidirectional CDW also exists in the Domain-I area, and it is parallel with the chain-like superstructure direction (Figures 2d and 2f). This makes the unidirectional CDW less visible in the STM topography taken on the Domain-I surface (Figure 2b). The d$I$/d$V$ spectra are similar on these two NiAs surfaces (see Figure S5), which is also consistent with the conclusion that the unidirectional CDW exists on both NiAs surfaces. We find that no matter what type of NiAs surface is, the unidirectional CDW on the top Ba layer always has the same direction as that on the underneath NiAs surface (see Figure S6).

After characterizing the triclinic $BaNi_2As_2$, we increase the temperature to be above the phase transition temperature (~136 K) and study the tetragonal $BaNi_2As_2$. Figures 3a-3c show the STM topographies on the Domain-II NiAs surface taken at 120 K, 140 K and 150 K, respectively, and Figures 3d-3f are their FT images (see Figure S7). As increasing the temperature to be above the phase transition temperature (~136 K), the $Q = 1/3$ unidirectional CDW disappears on the NiAs surface (Figures 3d and 3e). Figures 3e and 3f also indicate that in the tetragonal phase, the NiAs surface becomes more ordered and the atomic-chain superstructure with the $1 \times 2$ periodicity starts to emerge at 150 K (see Figure S8). In the ordered superstructures, the two nearest As atomic chains have slightly different height. For the Ba surface, although the $Q = 1/3$ unidirectional CDW also disappears at 140 K and 150 K, no clear superstructure feature can be seen (Figures 3g and 3h, and see Figure S9). We also note that near the tetragonal to triclinic transition temperature, the previously reported incommensurate CDW does not appear in the STM topographies taken on the NiAs and Ba surfaces (Figures 3b and 3g). One possible reason could be that the incommensurate CDW is disturbed and weakened by the electric field from the STM tip. Similar effect has been observed in the impurity-pinned incommensurate CDW in $2H$-$NbS_2$ system.[28]

Finally, we investigate the influence of Sr substitution in the $Ba_{1-x}Sr_xNi_2As_2$ superconductor. Figures 4a and 4b are the STM topographies taken on the Ba/Sr and NiAs surfaces of the triclinic $Ba_{0.5}Sr_{0.5}Ni_2As_2$ ($T_c$ ~ 1.3 K, see Figure S1), and the insets are their FT images. We can see that there is no clear unidirectional CDW feature on the Ba/Sr surface (Figure 4a). Although $Ba_{0.5}Sr_{0.5}Ni_2As_2$ is in the triclinic phase, both the complex chain-like structures and the unidirectional CDW disappear on the NiAs surface, and the NiAs surface appears as the atomic-chain superstructures with the $1 \times 2$ periodicity (Figure 4b, and see Figure S10). The NiAs surface shown in Figure 4b is similar as that for the tetragonal $BaNi_2As_2$ (Figure 3c), which indicates that the Sr substitution suppresses the unidirectional CDW and stabilizes the $1 \times 2$ superstructure on the NiAs surface of the triclinic $Ba_{0.5}Sr_{0.5}Ni_2As_2$. This could be due to the smaller size of the Sr atom, which makes the lattice constant along the $c$ direction in $Ba_{0.5}Sr_{0.5}Ni_2As_2$ smaller than that in $BaNi_2As_2$ (Fig. S1b). In this case, the Sr substitution could induce the chemical pressure effect that stabilizes the $1 \times 2$ superstructure on the NiAs surface of $Ba_{0.5}Sr_{0.5}Ni_2As_2$. We also note that this $1 \times 2$ superstructure is a structural modulation rather than a CDW order, which is different as the previously reported $Q = 1/2$ CDW in the $Ba_{1-x}Sr_xNi_2As_2$ system.[10]

Figure 4c shows the comparison between the d$I$/d$V$ spectra taken on the Ba surface of $BaNi_2As_2$ and the Ba/Sr surface of $Ba_{0.5}Sr_{0.5}Ni_2As_2$, where we can see the peaks related with the bottom of the α band are located at similar energy position ($\sim -150$ mV). These data indicate that substituting Ba with Sr in $BaNi_2As_2$ can suppress the unidirectional CDW without inducing the electronic doping effect. Then we measure the superconducting gaps in $BaNi_2As_2$ and $Ba_{0.5}Sr_{0.5}Ni_2As_2$ at ~75 mK. As shown in Figure 4d, the superconducting coherence peaks for $BaNi_2As_2$ and $Ba_{0.5}Sr_{0.5}Ni_2As_2$ are located at ~90 μV and ~125 μV, respectively (see Figure S11). This indicates that the size of the superconducting gap in $Ba_{0.5}Sr_{0.5}Ni_2As_2$ is ~39% larger than that in $BaNi_2As_2$. Figures 4e and 4f show the vortices taken on the $BaNi_2As_2$ and $Ba_{0.5}Sr_{0.5}Ni_2As_2$, respectively. They are imaged by taking the d$I$/d$V$ maps inside the superconducting gap with a magnetic field applied perpendicular to the surface. From the linecut d$I$/d$V$ spectra shown in Figures 4g and 4h, we can see that the superconducting gaps are strongly suppressed in the center of the vortices. Interestingly, while the vortices in $Ba_{0.5}Sr_{0.5}Ni_2As_2$ are more or less isotropic, the vortices in the $BaNi_2As_2$ are elongated along the unidirectional CDW direction (see Figure S12), which may indicate that the electronic nematicity in $BaNi_2As_2$ is related with its unidirectional CDW.

In conclusion, in the triclinic $BaNi_2As_2$, we provide direct evidences for the unidirectional CDW with $Q = 1/3$ on both the Ba and NiAs surfaces. Due to the rebonding of the As anions, the complex chain-like superstructures appear on the NiAs surface. For the $Ba_{0.5}Sr_{0.5}Ni_2As_2$ sample, we find an ordered atomic-chain superstructure with $1 \times 2$ periodicity on the NiAs surface, and the unidirectional CDW is suppressed on both the Ba/Sr and NiAs surfaces. We also show that in comparison with $BaNi_2As_2$, the superconducting gap is enhanced in $Ba_{0.5}Sr_{0.5}Ni_2As_2$, which could be due to the suppression of unidirectional CDW and the stabilization of the $1 \times 2$ superstructure in $Ba_{0.5}Sr_{0.5}Ni_2As_2$. Our results provide detailed microscopic information for fully understanding the complex structure, unidirectional CDW, and superconductivity in this class of pnictide superconductors.

## MATERIALS AND METHODS

**Sample Synthesis and Characterization:** Single crystals of $Ba_{1-x}Sr_xNi_2As_2$ were synthesized by the self-flux method which is similar as the procedure described in Ref. [29]. The tetragonal to triclinic phase transition in $Ba_{1-x}Sr_xNi_2As_2$ can be seen in the temperature-dependent electrical resistivity measurements (see Figure S1).

**STM Measurements:** The low-temperature and the variable-temperature STM experiments were carried out with a Unisoku low-temperature STM. $Ba_{1-x}Sr_xNi_2As_2$ single crystal samples were cleaved at 77 K and then transferred into the STM head for measurements. Chemically etched tungsten tips were used for the STM measurements. STS measurements were done by using a standard lock-in technique.

## ASSOCIATED CONTENTS

### Supporting Information:

Temperature-dependent resistivity and x-ray diffraction spectra of $BaNi_2As_2$ and $Ba_{0.5}Sr_{0.5}Ni_2As_2$; Bias-voltage-dependent STM topographies taken on the Ba surface of $BaNi_2As_2$; Bias-voltage-dependent d$I$/d$V$ maps and the Fourier transform (FT) images on the Ba surface of $BaNi_2As_2$; Autocorrelation of the constant energy contours consisting of the α pockets in $BaNi_2As_2$; d$I$/d$V$ spectra taken on the Domain-I and Domain-II NiAs surfaces of $BaNi_2As_2$; STM topographies taken near the step edge between the Ba surface and the NiAs surface of $BaNi_2As_2$; Temperature-dependent STM topographies on the Domain-I NiAs surface of $BaNi_2As_2$; Bias-voltage-dependent STM topographies taken at 150 K on the NiAs surface of $BaNi_2As_2$; Temperature-dependent STM topographies on the Ba surface of $BaNi_2As_2$; Bias-voltage-dependent STM topographies taken on the Ba surface of $Ba_{0.5}Sr_{0.5}Ni_2As_2$; Superconducting spectra taken on $BaNi_2As_2$ and $Ba_{0.5}Sr_{0.5}Ni_2As_2$; Superconducting vortices in $BaNi_2As_2$ and $Ba_{0.5}Sr_{0.5}Ni_2As_2$.

## AUTHOR INFORMATION

### Corresponding Authors

Correspondence and requests for materials should be addressed to S.Y., Y. Q or C.W.

### Author Contributions

S.Y. and Y. Q. conceived the experiments. T.Q., R.Z., S.S. and C.W. carried out the STM experiments and experimental data analysis. W.C. and Y.Q. were responsible for sample growth. C.W. and S.Y. wrote the manuscript with input from all authors.

**Notes**

The authors declare no competing financial interests.

**ACKNOWLEDGEMENTS**

S.Y. acknowledges the financial support from the National Key R&D Program of China (Grant No. 2020YFA0309602) and the start-up funding from ShanghaiTech University. C.W. acknowledges the support from National Natural Science Foundation of China (Grant No. 12004250) and the Shanghai Sailing Program (Grant No. 20YF1430700). Y.Q. acknowledges the financial support from the National Key R&D Program of China (Grant No. 2018YFA0704300), the National Natural Science Foundation of China (Grant No. U1932217, 11974246, 12004252) and Shanghai Science and Technology Plan (Grant No. 21DZ2260400). The authors thank the support from Analytical Instrumentation Center (# SPST-AIC10112914), SPST, ShanghaiTech University.

**REFERENCES**

(1) Fradkin, E.; Kivelson, S. A.; Tranquada, J. M. Colloquium: Theory of intertwined orders in high temperature superconductors. *Reviews of Modern Physics* **2015,** *87* (2), 457-482.

(2) Davis, J. C. S.; Lee, D.-H. Concepts relating magnetic interactions, intertwined electronic orders, and strongly correlated superconductivity. *Proceedings of the National Academy of Sciences* **2013,** *110* (44), 17623-17630.

(3) Cao, Y.; Rodan-Legrain, D.; Park, J. M.; Yuan, N. F. Q.; Watanabe, K.; Taniguchi, T.; Fernandes, R. M.; Fu, L.; Jarillo-Herrero, P. Nematicity and competing orders in superconducting magic-angle graphene. *Science* **2021,** *372* (6539), 264-271.

(4) Ghiringhelli, G.; Le Tacon, M.; Minola, M.; Blanco-Canosa, S.; Mazzoli, C.; Brookes, N. B.; De Luca, G. M.; Frano, A.; Hawthorn, D. G.; He, F.; Loew, T.; Sala, M. M.; Peets, D. C.; Salluzzo, M.; Schierle, E.; Sutarto, R.; Sawatzky, G. A.; Weschke, E.; Keimer, B.; Braicovich, L. Long-Range Incommensurate Charge Fluctuations in $(Y,Nd)Ba_2Cu_3O_{6+x}$. *Science* **2012,** *337* (6096), 821-825.

(5) Comin, R.; Damascelli, A. Resonant X-Ray Scattering Studies of Charge Order in Cuprates. *Annual Review of Condensed Matter Physics* **2016,** *7* (1), 369-405.

(6) Chu, J.-H.; Kuo, H.-H.; Analytis, J. G.; Fisher, I. R. Divergent Nematic Susceptibility in an Iron Arsenide Superconductor. *Science* **2012,** *337* (6095), 710-712.

(7) Kuo, H.-H.; Chu, J.-H.; Palmstrom, J. C.; Kivelson, S. A.; Fisher, I. R. Ubiquitous signatures of nematic quantum criticality in optimally doped Fe-based superconductors. *Science* **2016,** *352* (6288), 958-962.

(8) Fernandes, R. M.; Chubukov, A. V.; Schmalian, J. What drives nematic order in iron-based superconductors? *Nature Physics* **2014,** *10* (2), 97-104.

(9) Lee, S.; de la Peña, G.; Sun, S. X. L.; Mitrano, M.; Fang, Y.; Jang, H.; Lee, J.-S.; Eckberg, C.; Campbell, D.; Collini, J.; Paglione, J.; de Groot, F. M. F.; Abbamonte, P. Unconventional Charge Density Wave Order in the Pnictide Superconductor $Ba(Ni_{1-x}Co_x)_2As_2$. *Physical Review Letters* **2019,** *122* (14), 147601.

(10) Lee, S.; Collini, J.; Sun, S. X. L.; Mitrano, M.; Guo, X.; Eckberg, C.; Paglione, J.; Fradkin, E.; Abbamonte, P. Multiple Charge Density Waves and Superconductivity Nucleation at Antiphase Domain Walls in the Nematic Pnictide $Ba_{1-x}Sr_xNi_2As_2$. *Physical Review Letters* **2021,** *127* (2), 027602.

(11) Eckberg, C.; Campbell, D. J.; Metz, T.; Collini, J.; Hodovanets, H.; Drye, T.; Zavalij, P.; Christensen, M. H.; Fernandes, R. M.; Lee, S.; Abbamonte, P.; Lynn, J. W.; Paglione, J. Sixfold enhancement of superconductivity in a tunable electronic nematic system. *Nature Physics* **2020,** *16* (3), 346-350.

(12) Yao, Y.; Willa, R.; Lacmann, T.; Souliou, S.-M.; Frachet, M.; Willa, K.; Merz, M.; Weber, F.; Meingast, C.; Heid, R.; Haghighirad, A.-A.; Schmalian, J.; Le Tacon, M. An electronic nematic liquid in $BaNi_2As_2$. *Nature Communications* **2022,** *13* (1), 4535.

(13) Guo, Y.; Klemm, M.; Oh, J. S.; Xie, Y.; Lei, B.-H.; Gorovikov, S.; Pedersen, T.; Michiardi, M.; Zhdanovich, S.; Damascelli, A.; Denlinder, J.; Hashimoto, M.; Lu, D.; Mo, S.-K.; Moore, R. G.; Birgeneau, R. J.; Singh, D. J.; Dai, P.; Yi, M. Spectral Evidence for Unidirectional Charge Density Wave in Detwinned $BaNi_2As_2$. *arXiv* **2022,** *2205*, 14339v1. Submitted on 28 May 2022, DOI: 10.48550/arXiv.2205.14339

(14) Song, Y.; Wu, S.; Chen, X.; He, Y.; Uchiyama, H.; Li, B.; Cao, S.; Guo, J.; Cao, G.; Birgeneau, R. Phonon softening and slowing down of charge-density-wave fluctuations in $BaNi_2As_2$. *Physical Review B* **2023**, *107*, L04113.

(15) Souliou, S. M.; Lacmann, T.; Heid, R.; Meingast, C.; Paolasini, L.; Haghighirad, A. A.; Merz, M.; Bosak, A.; Tacon, M. L. Soft-phonon and charge-density-wave formation in nematic $BaNi_2As_2$. *Physical Review Letters* **2022**, *129*, 247602.

(16) Merz, M.; Wang, L.; Wolf, T.; Nagel, P.; Meingast, C.; Schuppler, S. Rotational symmetry breaking at the incommensurate charge-density-wave transition in $Ba(Ni, Co)_2(As, P)_2$: Possible nematic phase induced by charge/orbital fluctuations. *Physical Review B* **2021**, *104* (18), 184509.

(17) Pokharel, A. R.; Grigorev, V.; Mejas, A.; Dong, T.; Haghighirad, A. A.; Heid, R.; Yao, Y.; Merz, M.; Le Tacon, M.; Demsar, J. Dynamics of collective modes in an unconventional charge density wave system $BaNi_2As_2$. *Communications Physics* **2022,** *5* (1), 141.

(18) Ronning, F.; Kurita, N.; Bauer, E. D.; Scott, B. L.; Park, T.; Klimczuk, T.; Movshovich, R.; Thompson, J. D. The first order phase transition and superconductivity in $BaNi_2As_2$ single crystals. *Journal of Physics: Condensed Matter* **2008**, *20* (34), 342203.

(19) Subedi, A.; Singh, D. J. Density functional study of $BaNi_2As_2$: Electronic structure, phonons, and electron-phonon superconductivity. *Physical Review B* **2008**, *78* (13), 132511.

(20) Kurita, N.; Ronning, F.; Tokiwa, Y.; Bauer, E. D.; Subedi, A.; Singh, D. J.; Thompson, J. D.; Movshovich, R. Low-Temperature Magnetothermal Transport Investigation of a Ni-Based Superconductor $BaNi_2As_2$: Evidence for Fully Gapped Superconductivity. *Physical Review Letters* **2009,** *102* (14), 147004.

(21) Kudo, K.; Takasuga, M.; Okamoto, Y.; Hiroi, Z.; Nohara, M. Giant Phonon Softening and Enhancement of Superconductivity by Phosphorus Doping of $BaN_2As_2$. *Physical Review Letters* **2012,** *109* (9), 097002.

(22) Chen, Z. G.; Hu, W. Z.; Wang, N. L. Different nature of instabilities in $BaFe_2As_2$ and $BaNi_2As_2$ as revealed by optical spectroscopy. *physica status solidi (b)* **2010,** *247* (3), 495-499.

(23) Kothapalli, K.; Ronning, F.; Bauer, E. D.; Schultz, A. J.; Nakotte, H. Single-crystal neutron diffraction studies on Ni-based metal-pnictide superconductor $BaNi_2As_2$. *Journal of Physics: Conference Series* **2010**, *251*, 012010.

(24) Zeljkovic, I.; Walkup, D.; Assaf, B. A.; Scipioni, K. L.; Sankar, R.; Chou, F.; Madhavan, V. Strain engineering Dirac surface states in heteroepitaxial topological crystalline insulator thin films. *Nature Nanotechnology* **2015**, *10* (10), 849-853.

(25) Zhou, B.; Xu, M.; Zhang, Y.; Xu, G.; He, C.; Yang, L. X.; Chen, F.; Xie, B. P.; Cui, X.-Y.; Arita, M.; Shimada, K.; Namatame, H.; Taniguchi, M.; Dai, X.; Feng, D. L. Electronic structure of $BaNi_2As_2$. *Physical Review B* **2011**, *83* (3), 035110.

(26) Lei, B.-H.; Guo, Y.; Xie, Y.; Dai, P.; Yi, M.; Singh, D. J. Complex structure due to As bonding and interplay with electronic structure in superconducting $BaNi_2As_2$. *Physical Review B* **2022**, *105* (14), 144505.

(27) Li, Q.; Lin, W.; Yan, J.; Chen, X.; Gianfrancesco, A. G.; Singh, D. J.; Mandrus, D.; Kalinin, S. V.; Pan, M. Bond competition and phase evolution on the $IrTe_2$ surface. *Nature Communications* **2014**, *5* (1), 5358.

(28) Wen, C.; Xie, Y.; Wu, Y.; Shen, S.; Kong, P.; Lian, H.; Li, J.; Xing, H.; Yan, S. Impurity-pinned incommensurate charge density wave and local phonon excitations in $2H$-$NbS_2$. *Physical Review B* **2020**, *101* (24), 241404.

(29) Sefat, A. S.; McGuire, M. A.; Jin, R.; Sales, B. C.; Mandrus, D.; Ronning, F.; Bauer, E. D.; Mozharivskyj, Y. Structure and anisotropic properties of $BaFe_{2-x}Ni_xAs_2$ (x=0, 1, and 2) single crystals. *Physical Review B* **2009**, *79* (9), 094508.

**Figures**

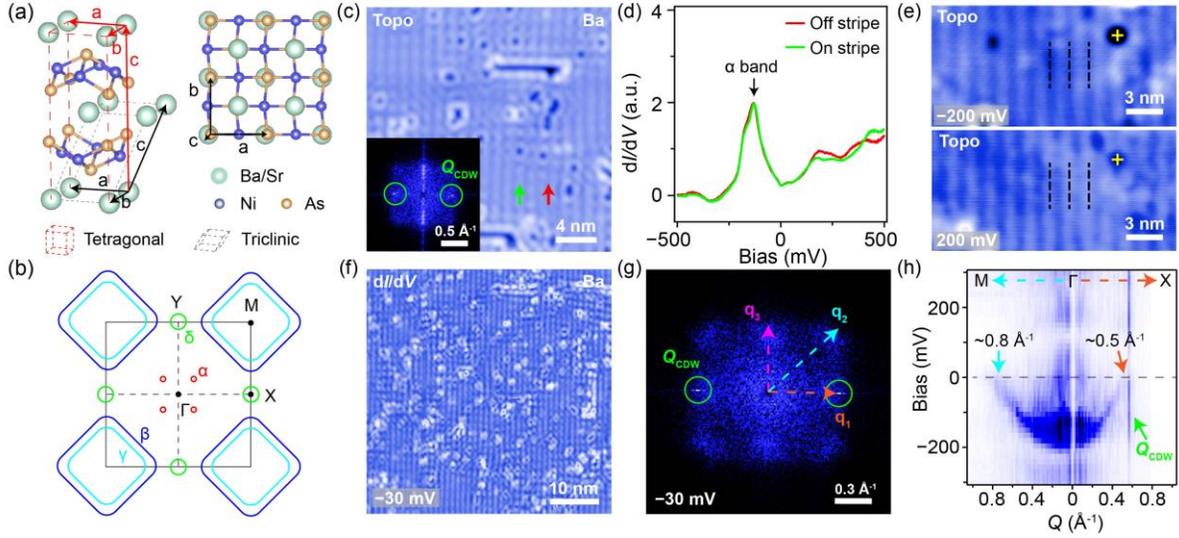

**Figure 1.** (a) Crystal structure of $Ba_{1-x}Sr_xNi_2As_2$ in the tetragonal (red lines) and triclinic (black lines) phases. (b) Schematic of the Fermi surface contours of the triclinic $BaNi_2As_2$ which are based on the previous ARPES data.[25] (c) Constant-current STM topography taken on the Ba surface of $BaNi_2As_2$ [$V_s = -25$ mV, $I = 1$ nA]. The inset shows its Fourier transform (FT) image. (d) Typical $dI/dV$ spectra taken on and off stripe marked in (c). (e) Constant-current STM topographies taken on the Ba surface of $BaNi_2As_2$ with $-200$ mV (upper panel) and 200 mV (lower panel) bias voltages. The yellow crosses mark the position of the same atomic defect. The black dashed lines show the phase shift of the stripe modulation between the two STM topographies. (f) $dI/dV$ map taken on the Ba surface with $-30$ mV bias voltage. (g) FT image of the $dI/dV$ map in (f). The colored arrows indicate the scattering vectors, and the green circles show the unidirectional CDW wavevector. (h) Linecuts along the directions of the $q_1$ and $q_2$ shown in (g) as a function of bias voltage. The STM data shown in this figure are taken at $T = 4.3$ K.

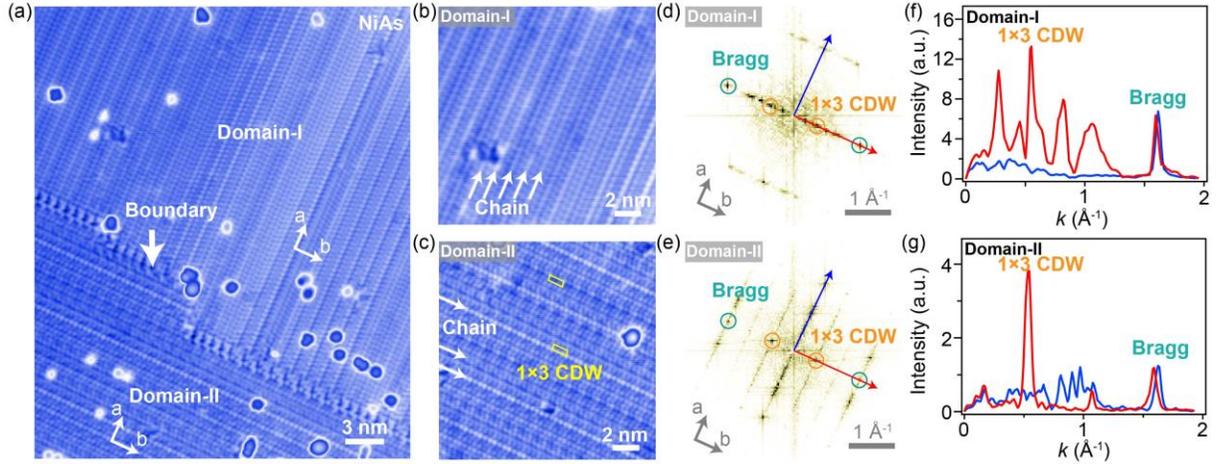

**Figure 2.** (a) Constant-current STM topography taken near the boundary between the Domain-I and Domain-II NiAs surfaces [$V_s = -100$ mV, $I = 20$ pA]. (b-c) Zoom-in STM topographies on the Domain-I and Domain-II NiAs surfaces, respectively [(b) $V_s = -100$ mV, $I = 20$ pA, (c) $V_s = -200$ mV, $I = 1$ nA]. The white arrows indicate the directions of the atomic chains. (d-e) FT images of the STM topographies in (b) and (c). (f) Linecut profiles along the red and blue arrows shown in (d). (g) Linecut profiles along the red and blue arrows in (e). The data shown in this figure are taken at $T = 4.3$ K.

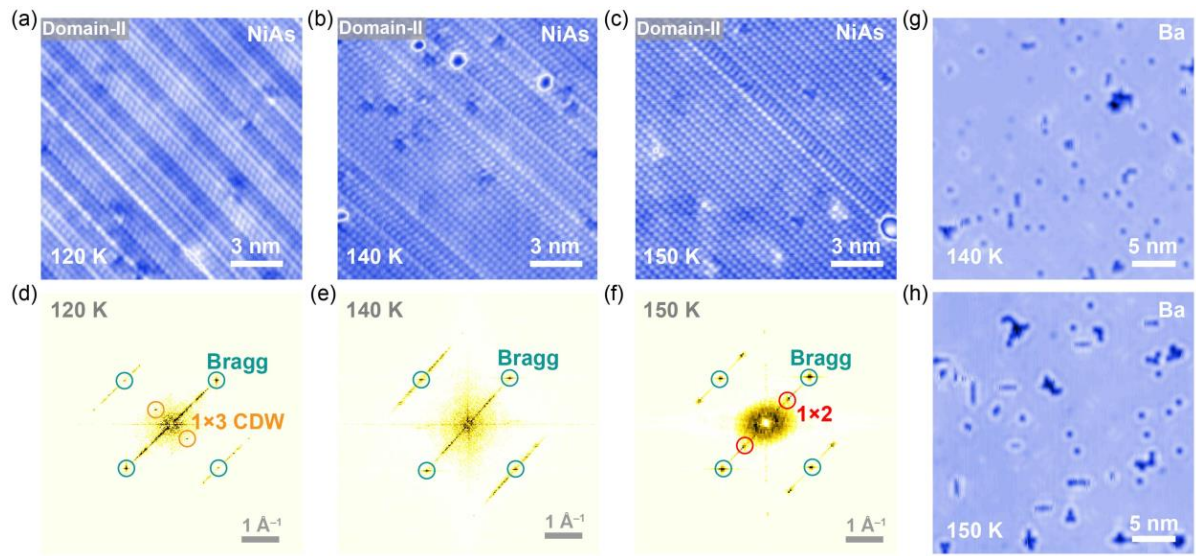

**Figure 3.** (a-c) Constant-current STM topographies taken on the Domain-II NiAs surface at 120 K [(a) $V_s = -100$ mV, $I = 50$ pA], 140 K [(b) $V_s = 100$ mV, $I = 100$ pA], and 150 K [(c) $V_s = -500$ mV, $I = 100$ pA]. (d-f) FT images of the STM topographies in (a-c). (g-h) Constant-current STM topographies taken on the Ba surface at 140 K [(g) $V_s = -50$ mV, $I = 100$ pA] and 150 K [(h) $V_s = -50$ mV, $I = 500$ pA], respectively.

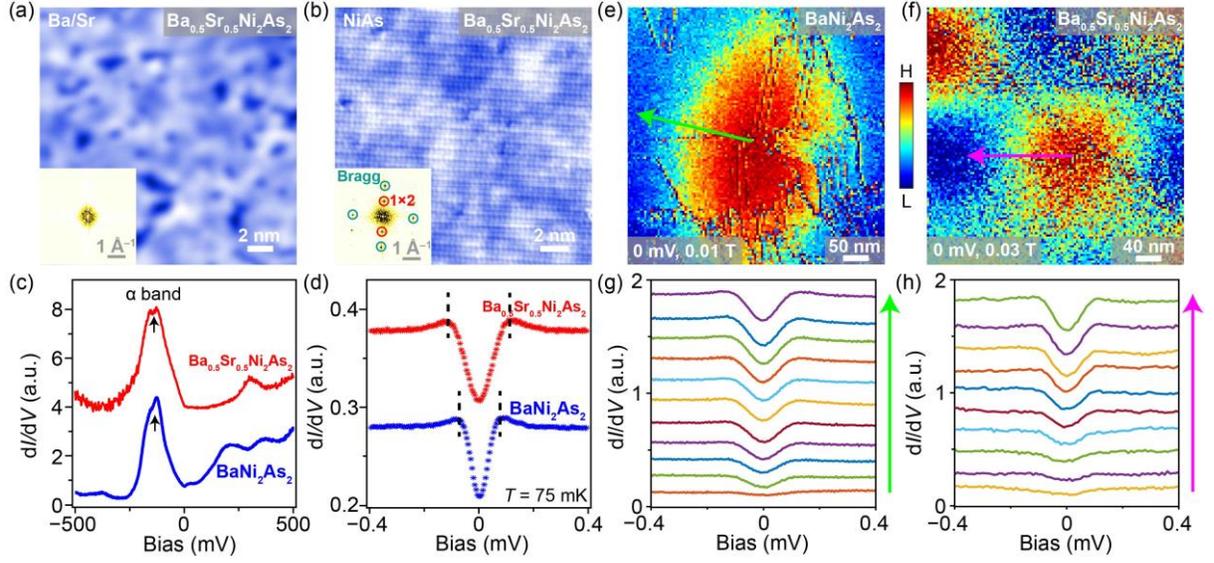

**Figure 4.** (a, b) Constant-current STM topographies taken on the Ba/Sr and NiAs surfaces of Ba$_{0.5}$Sr$_{0.5}$Ni$_2$As$_2$ at 4.3 K [(a) $V_s = -500$ mV, $I = 100$ pA, (b) $V_s = 50$ mV, $I = 100$ pA]. The insets are their FT images. (c) Typical d$I$/d$V$ spectra taken on the Ba/Sr surface of Ba$_{0.5}$Sr$_{0.5}$Ni$_2$As$_2$ (red) and the Ba surface of BaNi$_2$As$_2$ (blue). The spectra are vertically offset for clarity. (d) Typical superconducting spectra taken on the Ba/Sr surface of Ba$_{0.5}$Sr$_{0.5}$Ni$_2$As$_2$ (red) and the Ba surface of BaNi$_2$As$_2$ (blue). The spectra are vertically offset for clarity. (e, f) d$I$/d$V$ maps showing the superconducting vortices on BaNi$_2$As$_2$ and Ba$_{0.5}$Sr$_{0.5}$Ni$_2$As$_2$, respectively. (g, h) Spatially resolved d$I$/d$V$ spectra along the colored arrows shown in (e, f). The spectra are vertically offset for clarity. The data in (e-h) are taken at $T = 75$ mK.